\documentclass[a4paper]{jpconf}
\usepackage{graphicx}
\begin{document}
\title{Probing the Energy Structure of Positronium with a 203 GHz Fabry-Perot Cavity}

\author{T Suehara$^1$, A Miyazaki$^2$, A Ishida$^2$, T Namba$^1$, S Asai$^2$, T Kobayashi$^1$,
	H Saito$^3$, M Yoshida$^4$, T Idehara$^5$, I Ogawa$^5$, S Kobayashi$^5$, Y Urushizaki$^5$ and S Sabchevski$^6$}

\address{$^1$ International Center for Elementary Particle Physics (ICEPP), The University of Tokyo, 7-3-1 Hongo, Bunkyo-ku, Tokyo, 113-0033, Japan}
\address{$^2$ Department of Physics, Graduate School of Science, The University of Tokyo, 7-3-1 Hongo, Bunkyo-ku, Tokyo, 133-0033, Japan}
\address{$^3$ Department of General Systems Studies, Graduate School of Arts and Sciences, The University of Tokyo, 3-8-1 Komaba, Meguro-ku, Tokyo, 153-8902, Japan}
\address{$^4$ Accelerator Laboratory, High Energy Accelerator Research Organization (KEK), 1-1 Oho, Tsukuba, Ibaraki, 305-0801, Japan}
\address{$^5$ Research Center for Development of Far-Infrared Region, University of Fukui (FIR-FU), 3-9-1 Bunkyo, Fukui, Fukui, 910-8507, Japan}
\address{$^6$ Bulgarian Academy of Science, 1, 15 Noemvri Str., 1040 Sofia, Bulgaria}

\ead{suehara@icepp.s.u-tokyo.ac.jp}

\begin{abstract}
Positronium is an ideal system for the research of the bound state QED. 
The hyperfine splitting of positronium (Ps-HFS: about 203 GHz) is sensitive
to new physics beyond the Standard Model via a vacuum oscillation between an ortho-Ps and a virtual photon.
Previous experimental results of the Ps-HFS show 3.9\,$\sigma$ (15 ppm) discrepancy from the QED calculation.
All previous experiments used an indirect method with static magnetic field
to cause Zeeman splitting (a few GHz) between triplet states of ortho-Ps,
from which the HFS value was derived.
One possible systematic error source of the indirect method is the static magnetic field.
We are developing a new direct measurement system of the Ps-HFS
without static magnetic field.
In this measurement we use a gyrotron, a novel sub-THz light source,
with a high-finesse Fabry-Perot cavity to obtain enough radiation power at 203 GHz.
The present status of the optimization studies and current design of the
experiment are described.
\end{abstract}

\section{Introduction}

Positronium (Ps), the electron-positron bound state, is a purely leptonic system.
The energy difference between ortho-positronium (o-Ps, $1^3S_1$ state) and
para-positronium (p-Ps, $1^1S_0$ state), hyperfine splitting of positronium (Ps-HFS),
is a good target for precise bound state QED verification.
The Ps-HFS value is approximately 203 GHz (0.84 meV), which is significantly larger than hydrogen HFS (1.4 GHz).
One reason for the large Ps-HFS is the quantum oscillation:
o-Ps $\rightarrow \gamma^\ast \rightarrow$ o-Ps (o-Ps has the same quantum number as a photon).
Since some hypothetical particles (such as an axion or a millicharged particle) can
participate in the quantum oscillation, resulting in a shift of Ps-HFS value,
its precise measurement provides a probe for new physics beyond the Standard Model.

Measurements of the Ps-HFS have been performed in 70's and 80's
with the combined precision of 3.3 ppm\cite{Mills, Ritter},
The results were consistent with 1st order calculation of the QED available at that time.
However, the 2nd and 3rd order corrections have been calculated recently,
with the new prediction of 203.39169(41) GHz\cite{Kniehl}, 
deriving from the measured value of 203.38865(67) GHz by 3.9$\sigma$.
This discrepancy may come from new physics or the common systematic errors in the previous measurements.

In all previous measurements, the Ps-HFS transition was not directly measured, since
203 GHz was too high frequency to produce and control.
They measured Zeeman splitting of o-Ps instead.
A static magnetic field makes Zeeman mixing between $m_z=0$ state of o-Ps and p-Ps,
the resultant energy level of $m_z=0$ state becomes higher than $m_z = \pm1$ (Zeeman splitting).
This Zeeman splitting, which is proportional to the HFS energy level,
is a few GHz frequency if $\sim$ 1 Tesla magnetic field is applied.
They applied uniform magnetic field in RF cavities where positronium was introduced,
causing Zeeman transition and resultant o-Ps to p-Ps transition.
However, uncertainty (especially non-uniformity) of the static magnetic field,
which is very sensitive to the measured HFS value, can be a large source of systematic error.

In contrast to the indirect method, we plan to directly measure the
HFS transition, which does not need a static magnetic field, and is thus
free from the systematic error from the field.
To cause the direct transition, a powerful 203 GHz radiation field is necessary.
We are developing sub-THz to THz light source called a gyrotron, and also a high-finesse
Fabry-Perot cavity to accumulate sub-THz photons for the direct HFS measurement.

\section{Gyrotron}

\begin{figure}[h]
\begin{center}
\begin{minipage}{14pc}
\includegraphics[height=14pc]{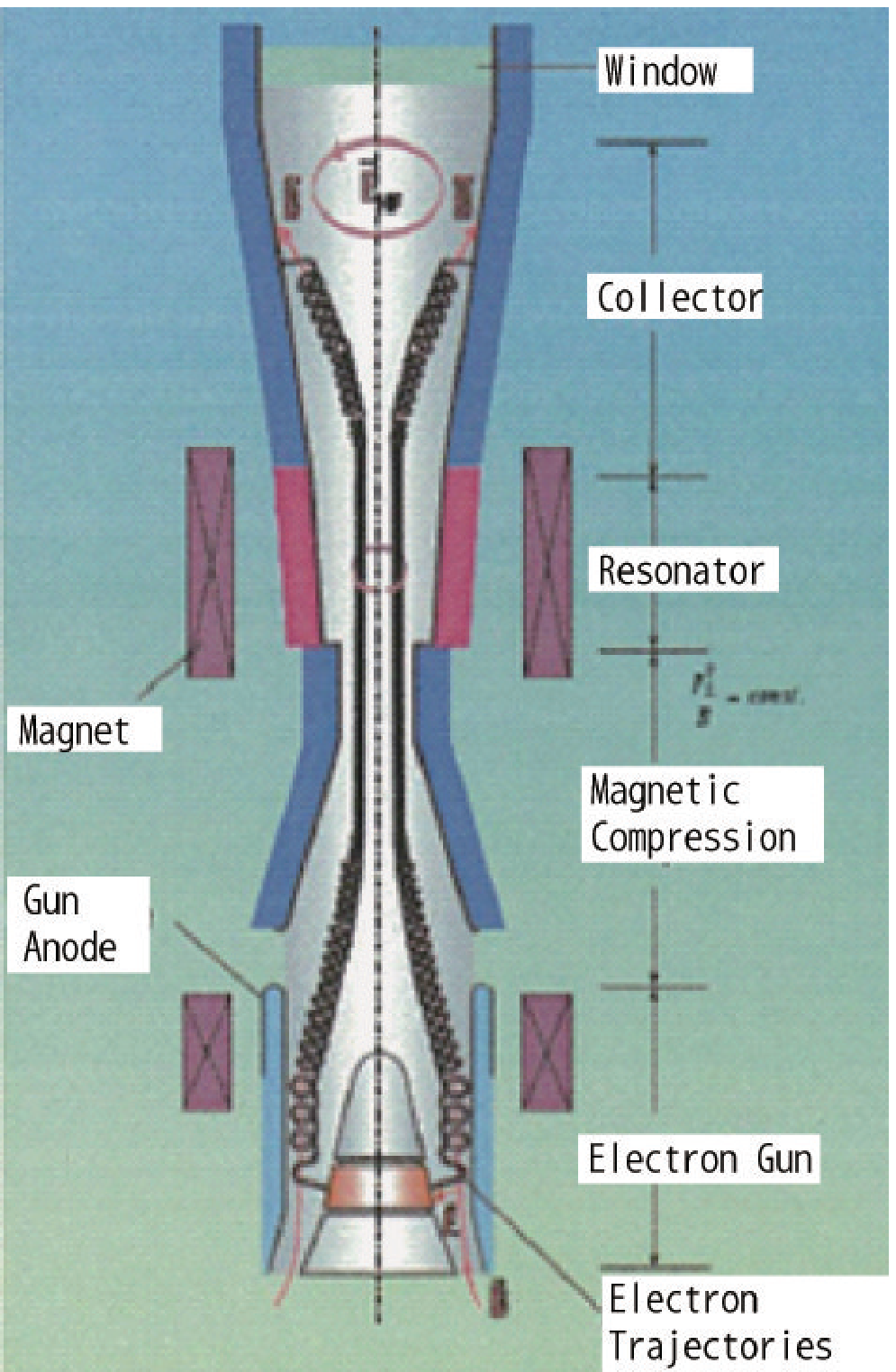}
\caption{\label{gyschematic}Schematic of gyrotrons.}
\end{minipage}\hspace{2pc}%
\begin{minipage}{14pc}
\includegraphics[height=14pc]{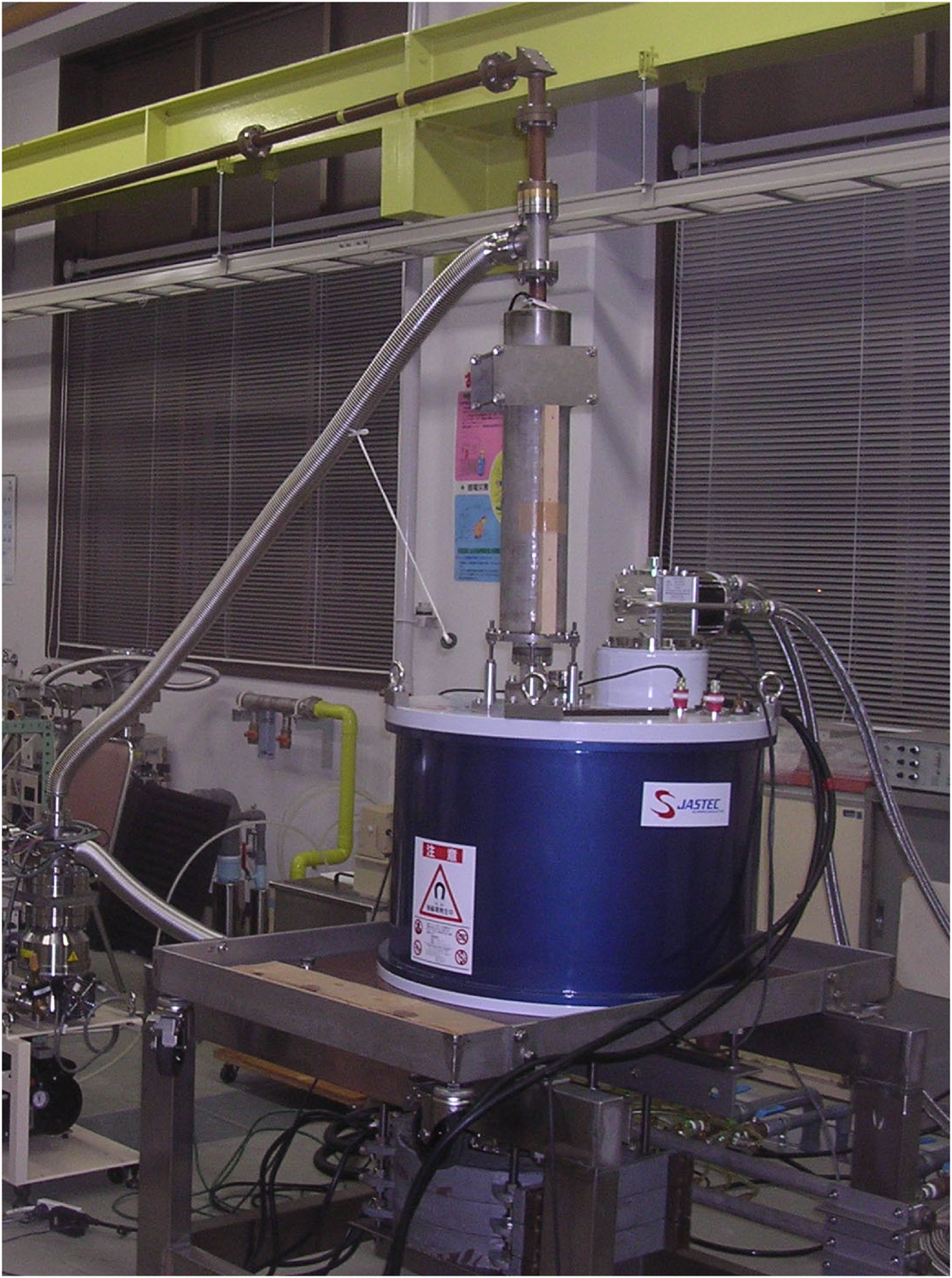}
\caption{\label{gypicture}Picture of the gyrotron for the HFS experiment.}
\end{minipage} 
\end{center}
\end{figure}

The gyrotron\cite{Idehara} is a novel high power radiation source for sub-THz to THz frequency region.
The structure of gyrotrons is shown in Figure \ref{gyschematic}.
The electrons are produced and accelerated at the DC electron gun, concentrated and rotated as cyclotron
motion in the superconducting magnet. The cyclotron frequency $f_c$ is
\begin{equation}
	f_c = \frac{eB}{2\pi{}m_0\gamma},
\end{equation}
where $B$ is the magnetic field strength, $m_0$ is the electron rest mass and
$\gamma$ is the relativistic factor of the electron.
A cavity is placed at the maximum magnetic field, whose resonance frequency is tuned just
to the cyclotron frequency.
The electrons stimulate resonance of the cavity and produce coherent photons at the cavity.
The photons are guided to the output port through the window, while electrons are dumped
at the collector.

We developed a gyrotron operating at $f_c = 203$ GHz with $B=7.425$ Tesla, $\gamma \sim 1.02$.
Figure \ref{gypicture} shows the gyrotron. The obtained radiation power is 609 W at the window,
which is reduced to 440 W during transmission through the waveguide system to the positronium cavity.
The spectral width is determined by $B$ uniformity and $\gamma$ spread by thermal distribution of
electrons, and is expected to be less than MHz, which is narrow enough to make resonance
at the Fabry-Perot cavity. Measurement results of a similar gyrotron shows
the spectral width is less than 10 kHz\cite{gywidth}.
The frequency can be tuned by changing the $\gamma$ factor with different acceleration of electrons,
but the tuning range is limited by the resonant width of the cavity to several handreds of megahertz.
Off-203 GHz radiation can be obtained by using different resonance mode of the cavity.
In our gyrotron, nearest resonant frequency to 203 GHz is at around 199 GHz.
The radiation power of the gyrotron will be monitored to account for corrections
to the transition probability.

\section{Fabry-Perot cavity}

Photons produced at the gyrotron are transported and accumulated in a cavity to
cause the Ps-HFS transition. Since 203 GHz ($\lambda = 1.475$ mm) photons can be treated optically
at the centimeter (or larger) size scale, we plan to use a Fabry-Perot cavity.
It consists of two opposing mirrors to confine photons between them.
Unlike RF cavities, the confinement in the Fabry-Perot cavity is 1-dimensional
while the other four sides are open.
We use a metal-mesh mirror on the input side of the cavity and
use a copper concave mirror on the other side.

\begin{table}[h]
\caption{\label{meshparam}
Mesh parameters with simulation and measurement result.
reflectance and transmittance are from estimations by the simulation,
and finesse is from results of the measurements.
The mesh of the last row will be used in the real experiment.
}
\begin{center}
\begin{tabular}{rrrrrr}
\br
mesh material & line width & line separation & reflectance & transmittance & finesse\\
\mr
gold   &  20 $\mu$m &  50 $\mu$m & 99.3\% & 0.32\% & 650\\
gold   &  10 $\mu$m &  50 $\mu$m & 98.6\% & 0.75\% & 290\\
silver &  50 $\mu$m & 130 $\mu$m & 96.9\% & 2.70\% & 180\\
silver & 400 $\mu$m & 240 $\mu$m & 99.2\% & 0.58\% & fabricating \\
\br
\end{tabular}
\end{center}
\end{table}

\begin{figure}[h]
\begin{center}
\begin{minipage}{11pc}
\includegraphics[width=11pc]{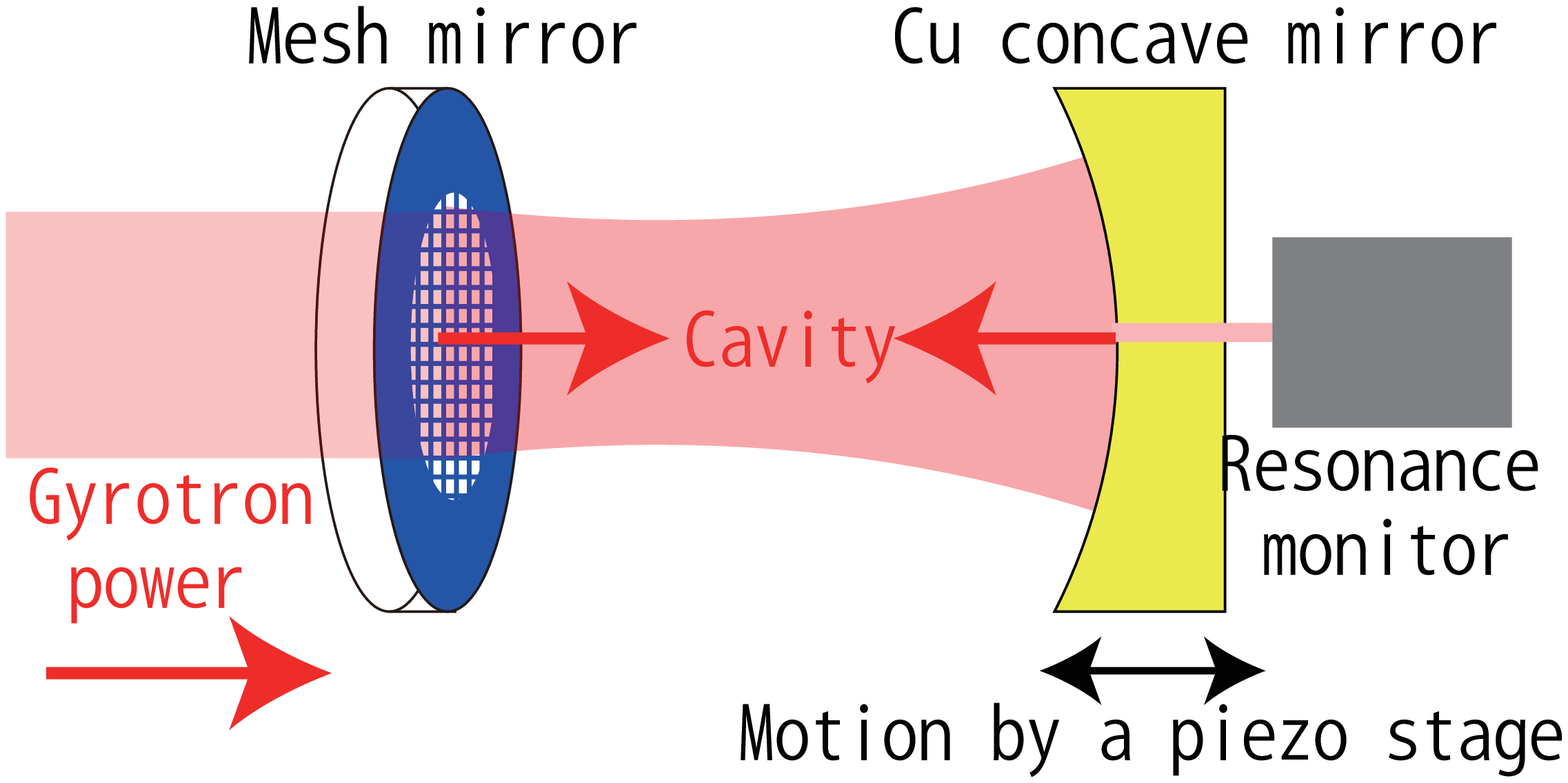}
\caption{\label{caschematic}Overview of the cavity.}
\end{minipage}\hspace{2pc}%
\begin{minipage}{11pc}
\includegraphics[width=11pc]{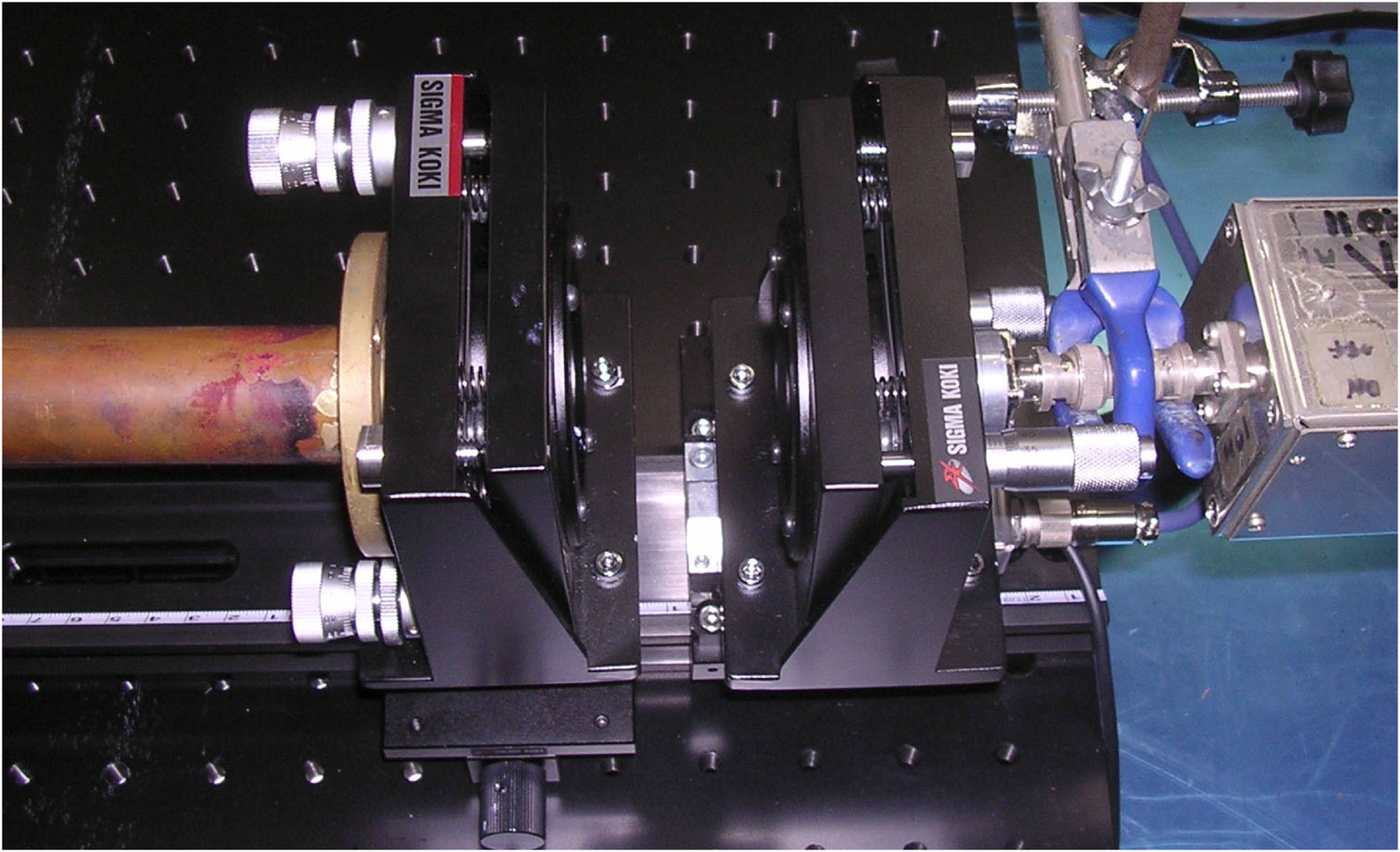}
\caption{\label{capicture}Picture of setup of the cavity test.}
\end{minipage} \hspace{2pc}
\begin{minipage}{11pc}
\includegraphics[width=11pc]{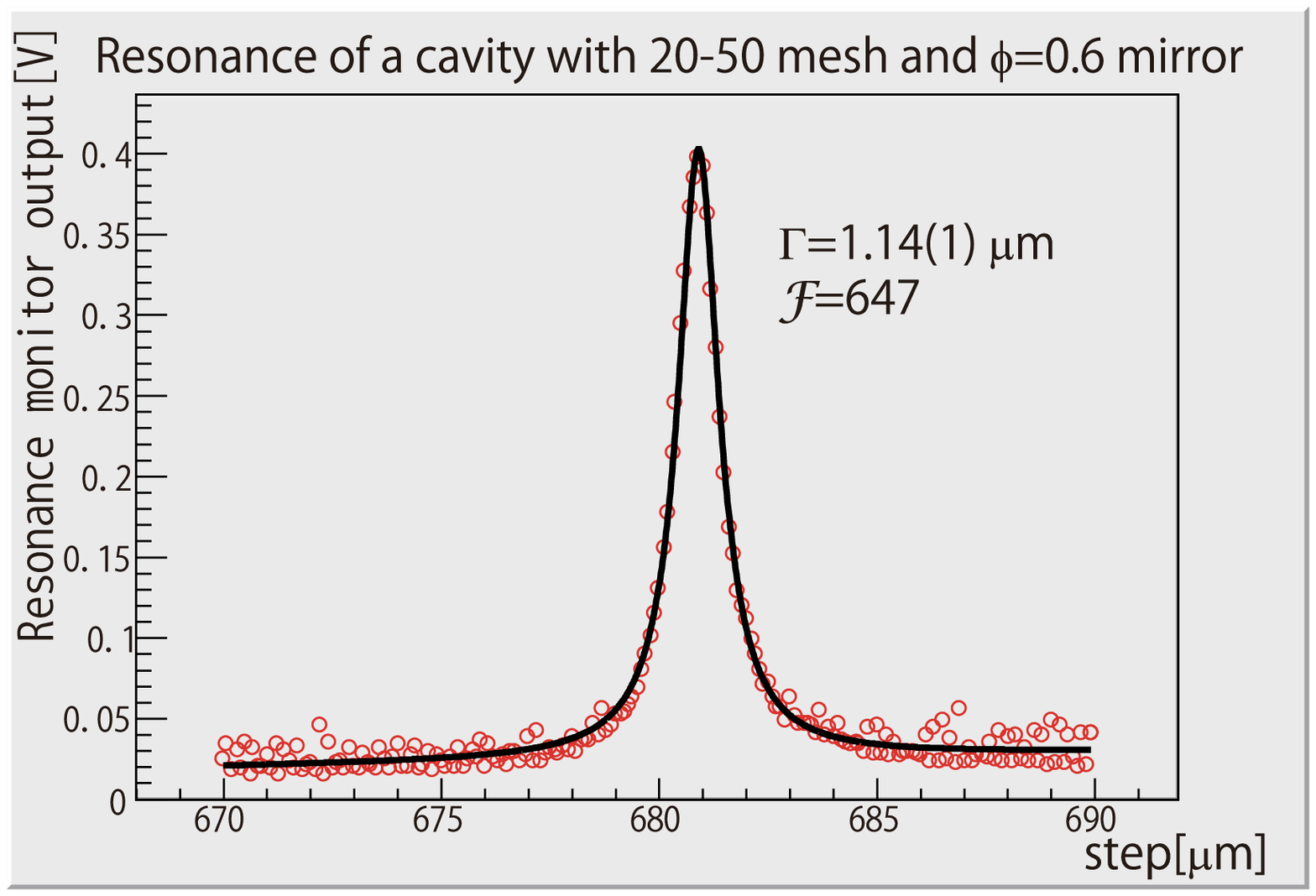}
\caption{\label{resonance}Resonance shape obtained by the test.}
\end{minipage} 
\end{center}
\end{figure}

The two most important characteristics of cavities are
``finesse" and ``input coupling".
Finesse can be written as $\mathcal{F} = 2\pi/(1-\rho)$,
where $\rho$ is the fraction of power left after one round-trip,
characterizing the capability of the cavity to store photons.
To maximize the finesse, power losses, which are categorized into
diffraction loss, medium loss and ohmic loss, must be minimized.
With the confinement of photons by the concave mirror and gas medium,
diffraction and medium loss are negligible in our cavity.
Ohmic loss occurs at the cavity mirror, which is around 0.15\% at the
copper mirror and more at the mesh mirror.
The ohmic loss at the mesh mirror varies by mesh parameters,
which can be calculated by field simulation (see Table \ref{meshparam}).
Input coupling is the fraction of input power introduced to the cavity mode,
an important parameter to introduce photons efficiently into the cavity.
In our cavity, the input coupling is mainly determined by
transmittance of the input mesh mirror, which is also calculated by the simulation.

The finesse and the input coupling were measured using various mesh and concave mirror
parameters. In the test setup, gyrotron power was introduced to the prototype cavity,
which consisted of a mesh mirror on the mirror mount and a concave mirror on a
mount on a piezo stage (see Figure \ref{caschematic} and \ref{capicture}).
Input, transmitted and reflected powers were monitored by three pyroelectric power monitors.
By shifting cavity length precisely by the piezo stage,
Breit-Wigner resonance shape was seen.
Finesse could be obtained from the width of the resonance with the transmit power monitor,
and the input coupling was seen by the reflection monitor.
With current best combination of the mesh and the concave mirror, $\mathcal{F} = 650$ was obtained (see Figure \ref{resonance}).
For the input coupling, concrete value could not be obtained because of interference of the reflection power
and difficulty to determine absolute power of the reflection/transmission because of non-optimal setup of the power measurements.
Input coupling of $\sim 20$\% can be estimated with the reflection data.

\section{Experimental Setup}

\begin{figure}[h]
\begin{center}
\includegraphics[height=14pc]{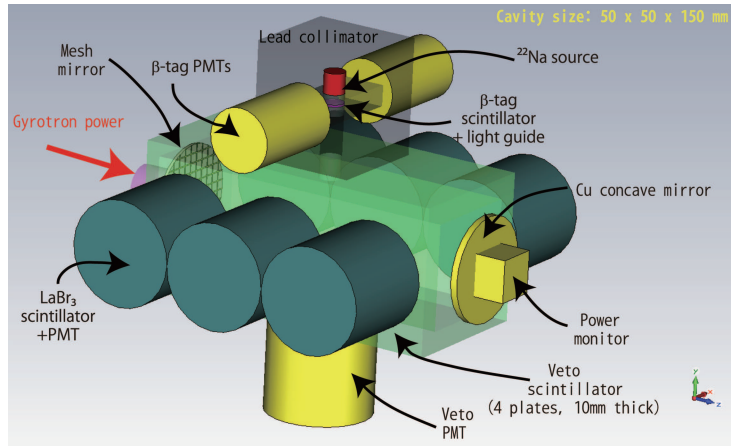}
\caption{\label{setup}Overview of the experimental setup.}
\end{center}
\end{figure}

\begin{table}[h]
\caption{\label{mcstat}Result of the Monte Carlo simulation.
Accidental bg.~does not include accidental tagging rate.
3 gamma decay is for the estimation of o-Ps background, and 2 gamma decay is for the estimation of signal rate.
3 gamma decay and 2 gamma decay does not include o-Ps production probability and
2 gamma decay does not include HFS transition probability.
For the 511 keV selection, an energy window of 20 keV (501 to 521 keV) is used.
Arrival time is not simulated.
For the final estimation of signal and background rate, see the text.
}
\begin{center}
\begin{tabular}{rrrrr}
\br
trigger       & all events      & accidental bg.      & 3 gamma decay & 2 gamma decay \\
              &                 & (non-tag, non-veto) & inside cavity & inside cavity \\
\mr
all           & $1.00\times10^7$& $7.36\times10^6$    & 28465         & 28500 \\
tagged        & $2.53\times10^6$& -                   & 25839         & 25930 \\
+non-veto     & $2.31\times10^6$& -                   & 18714         & 18981 \\
+single La    & 80848           & $1.60\times10^5$    & 7901          & 4967 \\
+511 keV      & 14350           & 9751                & 212           & 2186 \\
+double La    & 572             & 110                 & 75            & 417 \\
+511 keV both & 122             & 3                   & 6             & 114 \\
\br
\end{tabular}
\end{center}
\end{table}

Figure \ref{setup} shows a schematic view of the detection system.
Gyrotron power is introduced to the cavity via the mesh mirror, and accumulated between the mesh and the Cu mirror.
The sodium-22 source, which emits a positron with a 1275 keV photon, is located 40 mm above the cavity.
The emitted positrons pass through a $\beta$-tag scintillator
to generate start timing, a lead collimator (20 mm length, 15 mm aperture) and a hole in the veto scintillator
to reach the cavity.
LThe lead collimator also works as a shield to protect The LaBr$_3$ scintillator from 1275 and 511 keV (annihilation) photons emitted around the source.
The cavity is filled with mixed air of nitrogen and iso-C$_4$H$_{10}$ to form positronium atoms (20\% efficiency).
p-Ps (25\% of all Ps) annihilates to two 511 keV photons immediately as well as positron annihilation,
while o-Ps (75\%) remains with $\tau \sim 140$ ns and decays to three photons ($<$ 511 keV), generating delayed signal at LaBr$_3$
scintillator.
Six LaBr$_3$ scintillators surround the cavity to catch photons with energy resolution of $\sim 4$\% which can efficiently
separate 511 keV photons (evidence of HFS transition) from photons from o-Ps decay.
The LaBr$_3$ scintillators have timing resolution of $\sim 300$ psec to separate delayed events (signal)
from prompt events (annihilation or 1275 keV background).

The signal correction efficiency and background rate are estimated using Monte-Carlo simulation (GEANT4).
The major background processes are o-Ps contamination and accidental photons from positron creation and annihilation
without tag or veto signal which occur just around o-Ps lifetime from another event which is tagged but does not give energy deposit
on LaBr$_3$ scintillators.
Table \ref{mcstat} shows rates of the 2 gamma decay and each background with various trigger conditions.
Assuming 10 kW power accumulation at the cavity (50 W introduction from gyrotron, 200 times multiplication by the cavity),
HFS transition probability per o-Ps in the cavity is about 0.6\%.
With 1 MBq sodium-22 source, the signal rate is 0.01 Hz, including HFS transition probability, positronium formation rate of 20\%
in nitrogen and o-Ps ratio (75\%).
The accidental background is 0.014 Hz with accidental tagging rate of 4.6\% (200 ns timing window) and the o-Ps background is 0.09 Hz.
The S/N ratio is $0.01 / (0.014 + 0.09) = 0.096$, and significance of 30 $\sigma$ should be obtained in $10^6$ sec live time.
Background subtraction can be done by comparing with data without gyrotron power.

\section{Summary}

We are preparing the first experiment probing Ps-HFS with direct method to investigate discrepancy of Ps-HFS value
between theory and experiment.
We have developed a high power 203 GHz radiation source gyrotron and a Fabry-Perot cavity with a metal-mesh mirror
and a Cu concave mirror.
Monte-Carlo simulation of the detection system shows that the observation of Ps-HFS is feasible.
We are now assembling the detection system, and the first data taking is scheduled to start in this autumn.

\end{document}